\documentclass[aps,pra,twocolumn,eqsecnum,superscriptaddress,showpacs,floatfix,nofootinbib]{revtex4}

\usepackage{psfrag,graphicx}
\usepackage{dcolumn}
\usepackage{amsmath,amssymb}
\usepackage{bm}
\newcommand{\ra}{\rangle}

\newcommand{\ev}[1]{\mbox{$\langle #1 \rangle$}}

\newcommand{\ket}[1]{\left | \, #1 \right\rangle}

\newcommand{\be}{\begin{equation}}
\newcommand{\ee}{\end{equation}}
\newcommand{\bq}{\begin{eqnarray}}
\newcommand{\eq}{\end{eqnarray}}

\newcommand{\tr}{\text{tr}}

\newcommand{\unitmat}{\hbox{{\sf 1}\kern-.22em{\sf I}}}
\newcommand{\spinupket}{\ket{\! \uparrow}}
\newcommand{\spindownket}{\ket{\! \downarrow}}
\newcommand{\spinupbra}{\langle\uparrow\!\!|}
\newcommand{\spindownbra}{\langle\downarrow\!\!|}

\def\be{\begin{equation}}
\def\ee{\end{equation}}
\def\bea{\begin{eqnarray}}
\def\eea{\end{eqnarray}}

\def\tr{ \mbox{tr}}
\def\tr{{\rm tr}}
\def\1/2{\frac{1}{2}}

\begin{document}

\title{Quantum information and triangular optical
  lattices\footnote{Based on talk presented by J. K. Pachos at CEWQO 2004.}}

\author{Alastair Kay}
\affiliation{Department of Applied Mathematics and Theoretical
Physics, University of Cambridge, Cambridge CB3 0WA, UK,}
\author{Derek K. K. Lee}
\affiliation{Blackett
Laboratory, Imperial College, London SW7 2BW, UK,}
\author{Jiannis K. Pachos}
\affiliation{Department of Applied Mathematics and Theoretical
Physics, University of Cambridge, Cambridge CB3 0WA, UK,}
\author{Martin B. Plenio}
\affiliation{Blackett
Laboratory, Imperial College, London SW7 2BW, UK,}
\author{Moritz E. Reuter}
\affiliation{Blackett
Laboratory, Imperial College, London SW7 2BW, UK,}
\author{Enrique Rico}
\affiliation{Department d'Estructura i Constituents
de la Mat\` eria, Universitat de Barcelona, 08028, Barcelona, Spain.}

\date{\today}
\begin{abstract}

The regular structures obtained by optical lattice technology and
their behaviour are analysed from the quantum information
perspective. Initially, we demonstrate that a triangular optical
lattice of two atomic species, bosonic or fermionic, can be
employed to generate a variety of novel spin-$1/2$ models that
include effective three-spin interactions. Such interactions can
be employed to simulate specific one or two dimensional physical
systems that are of particular interest for their condensed matter
and entanglement properties. In particular, connections between
the scaling behaviour of entanglement and the entanglement
properties of closely spaced spins are drawn. Moreover, three-spin
interactions are well suited to support quantum computing without
the need to manipulate individual qubits. By employing Raman
transitions or the interaction of the atomic electric dipole
moment with magnetic field gradients, one can generate
Hamiltonians that can be used for the physical implementation of
geometrical or topological objects. This work serves as a review
article that also includes many new results.

\end{abstract}


\maketitle

\section{Introduction}

With the development of optical lattice technology
\cite{Raithel,Mandel1,Mandel3}, considerable attention has been
focused on the realisation of quantum computation \cite{Jaksch
99,Brennen 99,Kay 04,Mompart 03} as well as quantum
simulation of a variety of many-particle systems, such as spin chains
and lattices \cite{Jaks98,Kukl,Jask03,Duan}. This technology provides
the possibility to probe and realise complex quantum models with
unique properties in the laboratory. Examples that are of interest in
various areas of physics are systems that include many-body
interactions. These have been hard to study in
the past due to the difficulty in controlling them externally and
isolating them from the environment \cite{Mizel}. To overcome these
problems, techniques have been developed in quantum optics
\cite{Cirac1,Carl,Roberts} which minimise imperfections and
impurities in the implementation of the desired structures, thus
paving the way for the consideration of such ``higher order''
phenomena of multi-particle interactions. Their applications are
of much interest to cold atom technology as well as to condensed
matter physics and quantum information, some of which we shall see here.

The initial point of our study is the presentation of
the rich dynamics that governs the behaviour of an ultra cold atomic
ensemble when it is superposed with appropriate optical lattices.
For this purpose we consider the case of two species of atoms, denoted
here by $\uparrow$ and $\downarrow$ (see \cite{Jaks98,Kukl,Duan}),
trapped in the potential minima of a periodic lattice. These species
can be two different hyperfine ground states of the same atom, coupled,
via an excited state, by a Raman transition. The system is brought
initially into the Mott insulator phase where the number of atoms at
each site of the lattice is well defined. By restricting to the
case of only one atom per site, it is possible to characterise the
system by pseudo-spin basis states provided by the internal ground
states of the atom. Interactions between atoms in different sites
are facilitated by virtual transitions. These are dictated by the
tunneling coupling, $J$, from one site to its neighbours and by
collisional couplings, $U$, that take place when two or more atoms
are within the same site. Eventually the evolution of the system can
be effectively described by a wide set of spin interactions with
coupling coefficients completely controlled by the tunneling and
collisional couplings. This gives rise to the consideration of several
applications that are mainly related with the three-spin
interactions simulated on the lattice, which is the main focus of the
present article.

In that spirit, implementation of quantum simulation, of different
physical models
can be realised, with ground states that present a rich structure,
such as multiple degeneracies and a variety of quantum phase
transitions \cite{Sachdev,Sachdev1,Pachos04}. Some of these multi-spin
interactions have been theoretically studied in the past in the
context of the hard rod boson \cite{Pens88,Iglo89,Fend}, using
self-duality symmetries \cite{Turb,Pens82}. Phase transitions
between the corresponding ground states have been analysed
\cite{Iglo87,Chris}. Subsequently, these phases may also be viewed
as possible phases of the initial system, in the Mott insulator,
where the behaviour of its ground state can be controlled at will
\cite{Laughlin}. In this context the so-called cluster Hamiltonian
is of considerable interest as its ground state exhibits unique
entanglement properties. In this article we review the status quo
of the existing analysis of these entanglement properties as well
as presenting new results that indicate interesting links between
the entanglement properties of closely spaced spins and the
scaling of entanglement with spin separation. 

To implement quantum
computing, one can take advantage of the three-spin interaction to
construct multi-qubit gates that eventually can lead to quantum
computation without the need to manipulate single qubits, referred
to as `global addressing'. We use a single qubit to localise
operations, meaning that one and two qubit gates in a typical
quantum computing scheme are replaced by two and three qubit gates
in this scheme, which is what makes a triangular lattice with
three-spin interactions such a natural environment for
implementation of this concept. This global addressing lifts the
stringent experimental requirement of single atom addressing for
performing quantum gates. Moreover, error correction can be
performed without the need to make measurements during the
computation.

The paper is organised as follows. In Section \ref{lattice}, we
present the physical system and the conditions required to obtain
three-body interactions. The effective three-spin Hamiltonians for
the case of bosonic or fermionic species of atoms in a system of
three sites on a lattice are given in Sec. \ref{eff}. In Sec.
\ref{complex} we introduce complex couplings by considering the
effect Raman transitions can have on the tunneling process and
generalised effective Hamiltonians are presented that do not
preserve the number of atoms of each species. In addition, the
electric dipole of the atoms is considered which, through
interaction with an external inhomogeneous field, can generate
chirality. In Sec. \ref{local} a variety of entanglement
properties of spin chains with three-spin interactions is
analysed, and the cluster Hamiltonian is presented along with
novel connections between the scaling behaviour of entanglement
and the entanglement properties of closely spaced spins. A global
addressing quantum computation model is presented in Sec.
\ref{qcomputation} and, finally, in Sec. \ref{conclusions}
concluding remarks are given.

\section{The Physical Model}
\label{lattice}

\begin{center}
\begin{figure}[!h]
\resizebox{!}{2.9cm} {\includegraphics{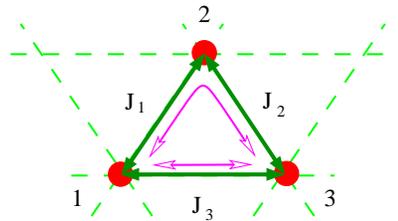} }
\caption{\label{first_triangle} The basic building block for the
    triangular lattice configuration. Three-spin
    interaction terms appear between sites $1$, $2$ and $3$. For
    example, tunneling between $1$ and $3$ can happen through two
    different paths, directly and through site $2$. The latter
    results in an exchange interaction between $1$ and $3$ that is
    influenced by the state of site $2$.}
\end{figure}
\end{center}

Let us consider a cloud of ultra cold neutral atoms superimposed
with several optical lattices
\cite{Jaks98,Kukl,Jask03,Duan,Pachos03}. For sufficiently strong
intensities of the laser field, this system can be placed in the
Mott insulator phase where the expectation value of only one
particle per lattice site is energetically allowed \cite{Mandel3}.
This still allows for the implementation of non-trivial manipulations
by virtual transitions that include energetically unfavourable states.
Here, we are particularly interested in the setup of lattices that form
an equilateral triangular configuration, as shown
in Fig. \ref{first_triangle}. This allows for the simultaneous
superposition of the positional wave functions of the atoms
belonging to the three sites. As we shall see in the following, this
results in the generation of a three-spin interaction.

The main contributions to the dynamics of the atoms in the lattice
sites are given by the collisions of the atoms within the same
site and the tunneling transitions of the atoms between
neighbouring sites. In particular, the coupling of the collisional
interaction for atoms in the same site are taken to be very large
in magnitude, while they are supposed to vanish when they are in
different sites. Due to the low temperature of the system, this
term is completely characterised by the s-wave scattering length.
Furthermore, the overlap of the Wannier wave functions between
adjacent sites determines the tunneling amplitude, $J$, of the
atoms from one site to its neighbours. Here, the relative rate
between the tunneling and the collisional interaction term is supposed
to be very small, i.e. $J\ll U$, so that the state of the system is
mainly dominated by the collisional interaction.

The Hamiltonian describing the three lattice sites with three
atoms of species $\sigma=\{\uparrow,\downarrow\}$ subject to the
above interactions is given by
\begin{equation}
H=H^{(0)}+V,
\label{ham}
\end{equation}
with
\begin{equation}
\begin{split}
H^{(0)}&=\frac{1}{2} \sum _{i \sigma \sigma'} U_{\sigma \sigma'}
a^{\dagger}_{i\sigma}a^{\dagger}_{i\sigma'}a_{i\sigma'}a_{i\sigma},
\\
V&=-\sum_{j\sigma} (J^\sigma_{j} a_{j\sigma}^\dagger a_{j+1
\sigma} +\text{H.c.}),
\end{split}
\nonumber
\end{equation}
where H.c. denotes the Hermitian conjugate, and $a_{j \sigma}$ denotes
the annihilation operator of atoms of
species $\sigma$ at site $j$. The annihilation operator can
describe fermions or bosons, satisfying commutation or
anticommutation relations respectively, given by
\begin{equation}
\begin{split}
&[a_{j \sigma},a_{j'\sigma'}^\dagger]_{\pm}=\delta_{jj'}
\delta_{\sigma \sigma' }, \\
&[a_{j\sigma},a_{j'\sigma'}]_{\pm}=[a_{j\sigma}^\dagger
,a_{j'\sigma'}^\dagger]_{\pm}=0,
\end{split}
\label{comm}
\end{equation}
where the $\pm$ sign denotes the anticommutator or the commutator.
The Hamiltonian $H^{(0)}$ is the lowest order in the expansion with
respect to the tunneling interaction.

Due to the large collisional couplings, activated when two or more
atoms are present within the same site, the weak tunneling
transitions do not change the average number of atoms per site.
This is achieved by adiabatic elimination of higher population
states during the evolution, leading to an effective
Hamiltonian \cite{Pachos 04}. The latter allows virtual transitions
between these levels providing eventually non-trivial
evolutions. According to this the low energy
evolution of the bosonic or fermionic system, up to the third order
in the tunneling interaction, is given by the effective Hamiltonian
\begin{equation}
H_{\text{eff}}=-\sum _\gamma {V_{\alpha \gamma} V_{\gamma \beta}
\over
  E_\gamma} +  \sum _{\gamma \delta} {V_{\alpha \gamma} V_{\gamma
    \delta} V_{\delta \beta} \over E_\gamma E_\delta} +{\cal O} ({J^4
  \over U^3}).
\label{effective}
\end{equation}
The indices $\alpha$, $\beta$ refer to states with one atom
per site while $\gamma$, $\delta$ refer to states with two or more
atomic populations per site, $E_\gamma$ are the eigenvalues of the
collisional part, $H^{(0)}$, while we neglected fast rotating
terms which is effective for long time intervals \cite{Pachos 04}.

It is instructive to estimate the energy scales involved in such a
physical system. We would like to have a significant effect of the
three-spin interaction within the decoherence
times of the experimental system, which we can take here to be of
the order of a few tens of ms. It is possible to vary the tunneling
interactions from zero to some maximum value which we can take
here to be of the order of $J/\hbar\sim$1 kHz \cite{Mandel1}. In
order to have a significant effect from the term $J^3/U^2$ within the
decoherence time, one should choose $U/ \hbar
\sim$10 kHz. This can be achieved experimentally by moving close to a
Feshbach resonance \cite{Inouye,Donley,Kokkelmans}, where
$U$ can take significantly large values. With
respect to these parameters we have $J/U\sim 10^{-1}$, which is
within the Mott insulator
regime, while the next order in perturbation theory is an order of
magnitude smaller than the one considered here and hence
negligible. This places the requirements of our proposal for
detecting the effect of three-spin interactions within the range
of the possible experimental values of the state of the art
technology.

\section{The effective three-spin Interactions}
\label{eff}

The perturbative dynamics of the system is better presented in terms
of effective spin interactions. Indeed, within the regime of single
atom occupancy per site, it is possible to switch to the pseudo-spin
basis of states of the site $j$ given
by $| \! \uparrow\rangle\equiv |n_{j\uparrow}=
1,n_{j\downarrow}=0\rangle$ and $|\! \downarrow\rangle \equiv
|n_{j\uparrow}= 0,n_{j\downarrow} = 1\rangle$. Hence, the
effective Hamiltonian can be given in terms of Pauli
matrices acting on states expressed in the pseudo-spin basis, as we
shall see in the following.

\subsection{The bosonic model}

In this subsection and the following one, we will develop the
Mott-Hubbard model in perturbation theory up to third order in $J/U$,
i.e. the ratio of the tunneling rate to the interaction
term. In the first case, we will study a bosonic system, when two atoms
of the same species are allowed to be in the same state. Eventually,
our model is described by
\begin{equation}
\label{ham1}
\begin{split}
H_{\text{eff}} &=\sum_{j=1}^3 \Big[A_j \mathbb{I} +  B_j
\sigma^z_j +
\\ &\lambda^{(1)}_j
\sigma^z_j \sigma^z_{j+1} + \lambda^{(2)}_j (\sigma^x_{j} \sigma^x_{j+1}
+\sigma^y_{j} \sigma^y_{j+1}) +
  \\ \\ &\lambda^{(3)} \sigma^z_{j} \sigma^z_{j+1}
  \sigma^z_{j+2}+\lambda^{(4)}_j (\sigma^x_{j} \sigma^z_{j+1}
  \sigma^x_{j+2}+ \sigma^y_{j} \sigma^z_{j+1} \sigma^y_{j+2})
  \Big],
\end{split}
\end{equation}
where $\sigma_j^{\alpha}$ is the $\alpha$ Pauli matrix at the site
$j$. The couplings $A$, $B$, and $\lambda^{(i)}$, are given in terms
of ${J^\sigma}/U_{\sigma \sigma'}$ by
\begin{equation}
\begin{split}
A_j=&-J_{1}^{\uparrow}J_{2}^{\uparrow}J_{3}^{\uparrow} \big(
\frac{9}{2U_{\uparrow \uparrow}^2}+ \frac{3}{2U_{\uparrow
\downarrow}^2} + \frac{3}{U_{\uparrow \downarrow}U_{ \uparrow
\uparrow }} \big)-
\\ & {J_{j}^{\uparrow}}^2\big(\frac{1}{U_{\uparrow \uparrow}}+
\frac{1}{2 U_{\uparrow \downarrow}} \big) + (\uparrow \leftrightarrow
\downarrow), \\
B_j= & -\frac{{J_{j}^{\uparrow}}^2 +{J_{j+2}^{\uparrow}}^2}{U_{\uparrow
      \uparrow}}-\frac{J_{1}^{\uparrow} J_{2}^{\uparrow}
      J_{3}^{\uparrow}}{U_{\uparrow \uparrow}}\big( \frac{1}{U_{\uparrow
      \downarrow}}+\frac{9}{2 U_{\uparrow \uparrow}}\big)-\\
      &( \uparrow\leftrightarrow \downarrow),
\nonumber
\end{split}
\end{equation}
\vspace{-0.5cm}
\begin{equation}
\begin{split}
\lambda^{(1)}_j =&-J_{1}^{\uparrow}J_{2}^{\uparrow}J_{3}^{\uparrow}
\big(\frac{9}{2U_{\uparrow \uparrow}^2}- \frac{1}{2U_{\uparrow
\downarrow}^2} - \frac{1}{U_{\uparrow \downarrow }U_{ \uparrow
\uparrow}} \big) -\\
& {J^{\uparrow}_{j}}^2 \big( \frac{1}{U_{\uparrow \uparrow}}
-\frac{1}{2U_{\uparrow \downarrow }}\big) + (\uparrow
\leftrightarrow
\downarrow), \\
\lambda^{(2)}_j=& - J_{j}^{\downarrow} J_{j+1}^{\uparrow}
J_{j+2}^{\uparrow} \big( \frac{3}{2U_{\uparrow
\downarrow}^2}+\frac{1}{2 U_{\uparrow \uparrow}^2}+
\frac{1}{U_{\uparrow \downarrow} U_{\uparrow \uparrow }} \big )-\\
& \frac{J^{\uparrow}_{j} J^{\downarrow}_{j}}{2 U_{\uparrow
    \downarrow}} + (\uparrow \leftrightarrow \downarrow),\\
\lambda^{(3)}=& -\frac{J_{1}^{\uparrow} J_{2}^{\uparrow}
  J_{3}^{\uparrow}}{U_{ \uparrow \uparrow }} \big( \frac{3}{2U_{
  \uparrow \uparrow}}-\frac{1}{U_{ \uparrow \downarrow} }\big)-
  (\uparrow \leftrightarrow \downarrow) ,\\
\lambda^{(4)}_j=&-\frac{J_{j}^{\uparrow} J_{j+1}^{\uparrow}
  J_{j+2}^{\downarrow}}{U_{ \uparrow \uparrow }} \big(\frac{1}{2 U_{
  \uparrow \uparrow}}+ \frac{1}{U_{\uparrow \downarrow }} \big ) -
  (\uparrow \leftrightarrow \downarrow),
\label{effcouplings}
\end{split}
\end{equation}
where the symbol $(\uparrow \leftrightarrow \downarrow)$ denotes the
repeating of the same term as on its left, but with the $\uparrow$ and
$\downarrow$ indices interchanged.

\begin{center}
\begin{figure}[!ht]
\resizebox{!}{7.0cm}{\includegraphics{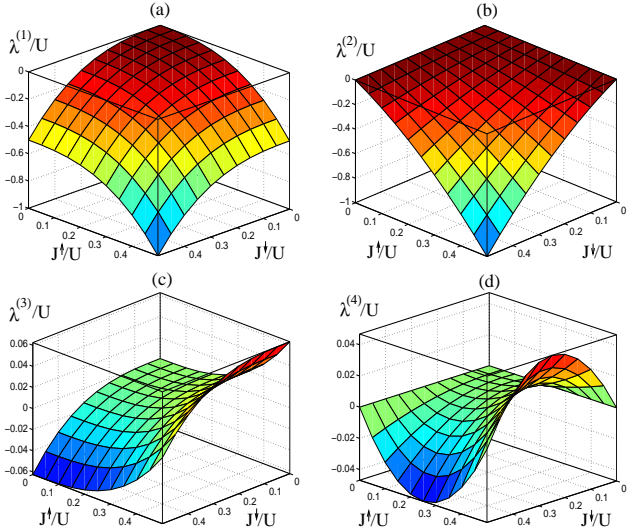}}
\caption{\label{comb1}The effective couplings (a) $\lambda^{(1)}$, (b)
  $\lambda^{(2)}$, (c) $\lambda^{(3)}$ and (d) $\lambda^{(4)}$ as
  functions of the tunneling couplings $J^{\uparrow}/U$ and
  $J^{\downarrow}/U$.  The tunneling couplings are set to be
  $J^\sigma_{1}=J^\sigma_{2} =J^\sigma_{3}$ and the collisional
  couplings to be $U_{ \uparrow \uparrow}=U_{\uparrow \downarrow
  }=U_{\downarrow \downarrow }=U$. All the parameters are normalised
  with respect to $U$.}
\end{figure}
\end{center}

\begin{center}
\begin{figure}[!ht]
\resizebox{!}{6.0cm}{\includegraphics{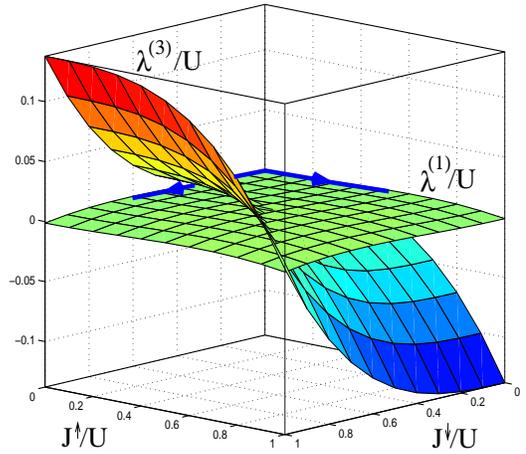}}
\caption{\label{zzz} The effective couplings $\lambda^{(1)}$ and
  $\lambda^{(3)}$ are plotted against $J^{\uparrow}/U$ and
  $J^{\downarrow}/U$ for $U_{ \uparrow \uparrow }=U_{\downarrow
    \downarrow }=2.12U$ and $U_{\uparrow \downarrow }=U$. The coupling
  $\lambda^{(1)}$ appears almost constant and zero as the unequal
  collisional terms can create a plateau area for a small range of the
  tunneling couplings, while $\lambda^{(3)}$ can be varied freely to
  positive or negative values.}
\end{figure}
\end{center}

Knowing the dependence of the effective couplings allows us to modify the
dynamics of the system at will, by changing the values of the tunneling
rate or coupling constant, as seen in Fig. \ref{comb1}. Moreover, one body
interactions in the Hamiltonian can be eliminated with an arbitrary
Zeeman term of the form $\sum_j \vec{B} \cdot \vec{\sigma_j}$ that can be
added by applying a Raman transition with the appropriate laser
fields. One can also isolate different parts of Hamiltonian
(\ref{ham1}), each one including a three-spin interaction term, by
varying the tunneling and/or the collisional couplings appropriately
so that particular terms in (\ref{ham1}) vanish, while others are
freely varied. An example of this can be seen in Fig. \ref{zzz} where
the couplings $\lambda^{(1)}$ and $\lambda^{(3)}$ are depicted. There,
for the special choice of the collisional terms, $ U_{\uparrow
  \uparrow }=U_{\downarrow \downarrow }=2.12U_{\uparrow  \downarrow
}$, the $\lambda^{(1)}$ coupling is kept to zero for a wide range of
the tunneling couplings, while the three-spin coupling,
$\lambda^{(3)}$, can take any arbitrary value. One can also suppress
the exchange interactions by keeping one of the two tunneling
couplings zero, without affecting the freedom in obtaining arbitrary
positive or negative values for $\lambda^{(3)}$, as seen in
Fig. \ref{zzz}.

\subsection{The fermionic model}

In this subsection, we will show the effective Hamiltonian for a
system where two atoms of the same species are not allowed to be in the
same site, so that they are described by  fermionic operators. For
the same reason, there is just one collisional coupling $U_{\uparrow
  \downarrow }=U$, and the others can be thought to contribute with an
infinite energy, $U_{ \uparrow \uparrow},U_{\downarrow
  \downarrow}\rightarrow \infty$. After a tedious calculation and keeping
terms to third order in $J_{i}^{\sigma} /U$, the effective Hamiltonian
appears as,

\begin{equation}
\begin{split}
&H_{\text{eff}}=\sum_{j=1}^3 \Big[ \mu^{(1)}_j (\mathbb{I}-
    \sigma_{j}^z \sigma_{j+1}^z) +\mu^{(3)}
(\sigma_{j}^z -\sigma_{1}^z \sigma_{2}^z \sigma_{3}^z)+\\
&\mu^{(2)}_{j} (\sigma_{j}^x \sigma_{j+1}^x \!\!+\!\sigma_{j}^y
    \sigma_{j+1}^y) \!+\!\mu^{(4)}_j (\sigma_{j}^x \sigma_{j+1}^z
    \sigma_{j+2}^x \!\!+\! \sigma_{j}^y \sigma_{j+1}^z \sigma_{j+2}^y
    ) \Big],
\end{split}
\nonumber
\end{equation}
where the effective couplings are functions of the tunneling and
collisional couplings, given by
\begin{equation}
\begin{split}
&\mu^{(1)}_j=
  -\frac{1}{2U}({J_{j}^{\uparrow}}^2 + {J_{j}^{\downarrow}}^2),
  \,\,\,\,\, \mu^{(2)}_{j}
  =\frac{1}{U}J_{j}^{\uparrow}J_{j}^{\downarrow},\\
&\mu^{(3)}=-\frac{1 }{2U^2}(J_{1}^{\uparrow} J_{2}^{\uparrow}
  J_{3}^{\uparrow}-J_{1}^{\downarrow} J_{2}^{\downarrow}
  J_{3}^{\downarrow} ),\\
&\mu^{(4)}_j=\frac{3 }{2U^2}(J_{j}^{\uparrow} J_{j+1}^{\uparrow}
  J_{j+2}^{\downarrow} - J_{j}^{\downarrow} J_{j+1}^{\downarrow}
  J_{j+2}^{\uparrow} ).
\end{split}
\nonumber
\end{equation}

\begin{center}
\begin{figure}[!ht]
\resizebox{!}{7.0cm}{\includegraphics{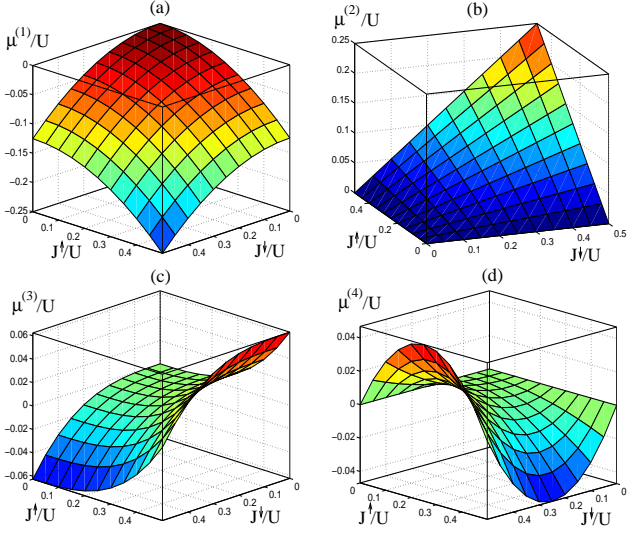}}
\caption{\label{comb2}The effective couplings (a) $\mu^{(1)}$, (b)
  $\mu^{(2)}$, (c) $\mu^{(3)}$ and (d) $\mu^{(4)}$ as functions of the
  tunneling couplings $J^{\uparrow}/U$ and $J^{\downarrow}/U$, where
  the tunneling couplings are set to be
  $J^\sigma_{1}=J^\sigma_{2}=J^\sigma_{3}$.}
\end{figure}
\end{center}

In this case, the dependence of the coupling
terms on the parameters of the initial Hamiltonian is simpler
than in the bosonic one. If the tunneling constants do not depend on
the pseudo-spin orientation, indicated here by the subscript
$j=1,2,3$, then any three-spin interaction
vanishes. Nevertheless, when the tunneling amplitudes depend on the
spin and there is just one of the orientation with non-zero tunneling,
then, only the two- and three-spin interactions in the $z$ direction
remain. A general picture of their behaviour can been seen in
Fig. \ref{comb2}.

\section{Complex couplings}\label{complex}

\subsection{Raman activated tunneling}

From the previous models, it is possible to create new, different,
Hamiltonians by employing techniques available from quantum optics
\cite{Jaks98,Duan,Liu}, like the application of Raman transitions
during the tunneling process. If the lasers producing the Raman
transition form standing waves, it is possible to activate
tunneling transitions of atoms that simultaneously experience a change
in their internal state. As we shall see in the following, the
resulting Hamiltonian is given by an $SU(2)$ rotation applied to each
Pauli matrix of the previous Hamiltonians.

Consider the case of activating the tunneling with the application of
two individual Raman transitions. These transitions consist of four
paired laser beams $L_1$, $L_2$ and $L_1'$, $L_2'$, each pair having a
blue de-tuning $\Delta$ and $\Delta'$, different for the two different
transitions. The phases and amplitudes of the laser beams can be
properly tuned so that the Raman transitions allow the tunneling of
the states
\begin{equation}
\begin{split}
&|+\rangle \equiv \cos \theta |a\rangle + \sin \theta e^{-i\phi } |b\rangle \\
&|-\rangle \equiv \sin \theta |a\rangle - \cos \theta e^{-i\phi } |b\rangle,
\nonumber
\end{split}
\end{equation}
or in a compact notation,
\be
\begin{pmatrix} |+\ra \\ |-\ra \end{pmatrix} = g(\phi,\theta)
\begin{pmatrix} |a\ra \\ |b\ra \end{pmatrix}
\ee
with the unitary $SU(2)$ matrix
\be
g(\phi,\theta) = \begin{pmatrix}   \cos \theta & e^{i \phi} \sin \theta \\
\sin \theta & -e^{i\phi} \cos \theta \end{pmatrix}
\ee
In the above equations, $\phi$ denotes the  phase difference between
the $L_i$ laser field, while $\tan \theta =|\Omega_2 /\Omega_1|$.
$\Omega_i$ are the Rabi frequencies of the laser fields. Hence, the
resulting tunneling Hamiltonian can be obtained from the initial one
via an $SU(2)$ rotation,
\be
V_c= gVg^\dagger =-\sum_i (J_+ {c^+_i}^\dagger c^+_{i+1}
+J_-{c^-_i}^\dagger c^-_{i+1} +\text{H.c.}).
\nonumber
\ee
where the
corresponding tunneling couplings are formally identified, i.e.
$J^+=J^\uparrow$ and $J^-=J^\downarrow$. Note that the collisional
Hamiltonian is not affected by the Raman transitions, and hence
it is not transformed under $g$ rotations.

\begin{center}
\begin{figure}[!ht]
\resizebox{!}{3.0cm}{\includegraphics{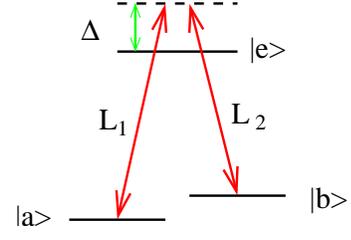}}
\caption{Example of a Raman transition activated by a pair of
  blue detuning laser fields $L_1$ and $L_2$.}
\end{figure}
\end{center}

It is easy to derive the effective Hamiltonian for this
transformation using the perturbative expansion. From
(\ref{effective}) we straightforwardly obtain the
second, third and, likewise, any order term of the
Hamiltonian $\tilde H_{\text{eff}}$ that appear after the
application of the Raman transition. They are given by an $SU(2)$
rotation that acts on the Pauli matrices of the initial effective
Hamiltonian and read
\begin{equation}
\tilde H^{(n)}_{\text{eff}}(\phi,\theta) = g(\phi,\theta)
H^{(n)}_{\text{eff}}g^\dagger (\phi,\theta), \nonumber
\end{equation}
where $n$ is the order of the perturbation.

This approach provides a variety of control parameters (e.g. the angle
$\phi$ and the ratio of the couplings of the two added Hamiltonians)
and, in addition, one can have these variables independent for each of
the three directions of the two dimensional (triangular) optical
lattice. Particular settings of these structures have been proved to
generate topological phenomena \cite{Duan} that support exotic
anionic excitations, useful for the construction of topological
memories \cite{Kitaev}. In addition, the possibility of varying,
arbitrarily, the control parameters gives us the natural setup to study
such phenomena as geometrical phases in lattice systems.

\subsection{Complex tunneling and topological effects}

Consider the case where we employ complex tunneling couplings
\cite{Zoller} to the optical lattice evolution. This can be
performed by employing additional characteristics of the atoms, like an
electric moment $\vec{d}_e$ and an external electromagnetic
field. As the external field can break time reversal symmetry, new
terms of the form $\{\sigma_{j}^x \sigma_{j+1}^y \sigma_{j+2}^z-
\sigma_{j}^y \sigma_{j+1}^x \sigma_{j+2}^z\}$ appear in the effective
Hamiltonian. In particular, the minimal coupling deduced from the
electric dipole of the atom
with the external field can be given, in general, by substituting
its momentum by
\begin{equation}
\vec{p} \rightarrow \vec{p} +(\vec{d}_e \cdot
\vec{\nabla})\vec{A}(\vec{x}), \nonumber
\end{equation}
where $\vec{A}$ is the corresponding vector potential. The new
term satisfies the
Gauss gauge if we demand that $\vec{\nabla} \cdot \vec{A}=0$, hence
it can generate a possible phase factor for the tunneling couplings.

\begin{center}
\begin{figure}[!ht]
\resizebox{!}{2.3cm} {\includegraphics{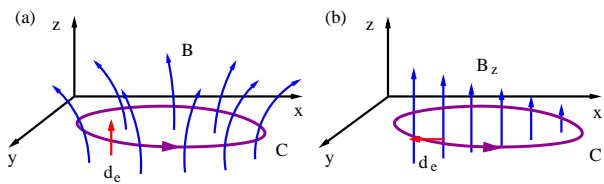} }
\caption{\label{zmag} The path circulation of the electric dipole
in the inhomogeneous magnetic field. Figure (a) depicts magnetic
field lines sufficient to produce the appropriate non-vanishing
gradient of $B_z$ along the $z$ axis. Figure (b) depicts only
$B_z$ and how it varies along the direction of the dipole,
$\vec{d}_e$.}
\end{figure}
\end{center}

Due to an evolution dictated by the differential form of the
Aharonov-Bohm effect \cite{Aharonov1} the cyclic move of an electric
moment through a gradient of a magnetic field contributes the
phase
\begin{equation}
\phi=\int_S(\vec{d}_e \cdot \vec{\nabla}) \vec{A} \cdot
d\vec{s},
\label{phase4}
\end{equation}
to the initial state, where $S$ is the surface enclosed by the cyclic
path of the
electric moment.
By inspection of relation (\ref{phase4}) we see that a nontrivial
phase can be produced if we generate an inhomogeneous magnetic
field in the neighbourhood of the dipole. In particular, a non-zero
gradient of the magnetic component perpendicular to the surface $S$,
varying in the direction of the dipole, ensures a non-zero phase
factor. For example, if we take $S$
to lie on the  $x$-$y$ plane and $\vec{d}_e$ is perpendicular to
the surface $S$, then a non-zero phase, $\phi$, is produced if
there is a non-vanishing gradient of the magnetic field along the
$z$ direction. This is sketched in Figure \ref{zmag}(a), where the
magnetic lines are plotted such that they produce the proper
variation of $B_z$ in the $z$ direction. Alternatively, if
$\vec{d}_e$ is along the surface plane, then a non-zero phase is
produced if the $z$ component of the magnetic field has a
non-vanishing gradient along the direction of $\vec{d}_e$ as seen
in Figure \ref{zmag}(b), where only the $z$ component of the
magnetic field has been depicted.
\vspace{-0.3cm}
\vspace{-0.7cm}

We can denote by $J=e^{i\phi}|J|$ the generation of the phase factor
of the tunneling coupling between two neighbouring sites, with
\begin{equation}
\phi=\int_{\vec{x}_i}^{\vec{x}_{i+1}}(\vec{d}_e \cdot
\vec{\nabla})\vec{A} \cdot d \vec{x}. \nonumber
\end{equation}
Here $\vec{x}_i$ and $\vec{x}_{i+1}$ denote the positions of the
lattice sites connected by the tunneling coupling $J$.

\begin{center}
\begin{figure}[!ht]
\resizebox{!}{2.9cm} {\includegraphics{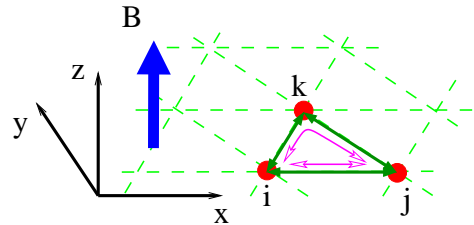} }
\caption{\label{triangle} The basic building block for the triangular
    lattice configuration. Exchange of atoms through sites $i$ and $j$
    can happen directly or through site $k$, but with a phase
    difference, causing the generation of the chiral three-spin
    interaction.}
\end{figure}
\end{center}

In order to isolate the new effects generated by the consideration
of complex tunneling couplings, we restrict ourselves to purely
imaginary ones, i.e. $J_i^{\sigma}= \pm i |J_i^{\sigma}|$. We
also focus, initially, on the case where the optical lattices
generate a two dimensional structure of equilateral triangles, as
in Figure \ref{triangle}. Such a non-bipartite structure is
necessary in order to manifest the breaking of the symmetry under
time reversal, $T$, in our model, eventually producing an effective
Hamiltonian that is not invariant under complex conjugation of the
tunneling couplings. Moreover, as the second order perturbation
theory is manifestly $T$ symmetric, we need to consider the third
order.
\vspace{-0.3cm}
\vspace{-0.7cm}

In this case, the effective Hamiltonian \cite{Pachos}, up to the second
order in perturbation theory, becomes
\begin{equation}
\begin{split}
H_{\text{eff}}^{(2)}=\!\sum_i \!\Big[&A
  \mathbb{I} + B \sigma^z_i +C \sigma^z_i\sigma^z_{i+1} \\
&+D(\sigma^x_i\sigma^x_{i+1}+ \sigma^y_i\sigma^y_{i+1})
\Big]
\nonumber
\end{split}
\end{equation}
The above couplings are given in perturbation theory by
\begin{equation}
\begin{split}
&A= -{{|J^\uparrow}|^2 \over U_{\uparrow\uparrow}}
-{{|J^\downarrow}|^2
  \over U_{\downarrow\downarrow}} -{|{J^\uparrow}|^2+|{J^\downarrow}|^2
\over 2 U_{\uparrow \downarrow}}
,\,\,\,
B= -{|{J^\uparrow}|^2 \over 2U_{\uparrow\uparrow}} +{|{J^\downarrow}|^2
  \over2 U_{\downarrow\downarrow}},\\
& C=-{|{J^\uparrow}|^2 \over U_{\uparrow\uparrow}}
-{|{J^\downarrow}|^2
  \over U_{\downarrow\downarrow}} +{|{J^\uparrow}|^2+|{J^\downarrow}|^2
\over 2 U_{\uparrow \downarrow}}
,\,\,\,
D={J^\uparrow {J^\downarrow}^*+{J^\uparrow}^* J^\downarrow \over
  2U_{\uparrow\downarrow}}. \nonumber
\end{split}
\end{equation}
It is easily seen that the values of the above interactions
remain unchanged compared to the case with no magnetic field. This
is the case since, up to the second order perturbation theory, there
are no contributions from circular paths that can experience the
magnetic field. Note that it is possible to apply Raman
transitions that exactly cancel the $\sigma^z$ Zeeman term. Also,
one can choose $U_{\uparrow \downarrow} \rightarrow \infty$
(possibly by Feshbach resonances) so that $C$ vanishes. In
addition, one of the collisional couplings can be chosen to be
attractive, satisfying e.g. $U_{\uparrow\uparrow} =
-U_{\downarrow\downarrow} = -U$, so that, for
${J^\uparrow}^2={J^\downarrow}^2$, the $A$ coupling vanishes. Such
attractive collisional couplings can be achieved by Feshbach
resonances \cite{Cornish}. Stability of the system is maintained
if the attractive couplings are activated adiabatically, while the
atomic ensemble is in the Mott insulator regime. Since attractive
couplings may eventually lead to pairing on the same lattice site,
this negative $U$ interaction can only be applied for a short period.

The third order in the perturbative expansion includes
interactions between neighbouring sites via the circular path
around the elementary triangle. This brings in the effect of the
magnetic field, contributing a phase factor. While new interaction
terms appear, the couplings $A$, $B$, $C$ and $D$ remain the same.
Considering the case of collisional and tunneling couplings, such
that $C=D=0$, the effective Hamiltonian becomes, up to global
Zeeman terms,
\begin{equation}
H_{\text{eff}}^{(3)}=\sum_{\langle ijk\rangle} \left[
E(\sigma^x_i\sigma^y_j-
\sigma^y_i\sigma^x_j)+ F\epsilon_{lmn}\sigma^l_i
\sigma^m_j \sigma^n_k \right] \nonumber
\end{equation}
where
\begin{equation}
\begin{split}
&E= i{{J^\uparrow} J^\downarrow \over
  2U^2} (J^\uparrow+J^\downarrow) ,\,\,\,
F= i{{J^\uparrow} J^\downarrow \over
  2 U^2}(J^\uparrow-J^\downarrow).
\nonumber
\end{split}
\end{equation}
Here $\langle ijk\rangle$ denotes nearest neighbours,
$\epsilon_{lmn}$ with $\{l,m,n\}=\{x,y,z\}$ denotes the total
antisymmetric tensor in three dimensions and summation over the
indices $l,m,n$ is implied. From the expression of the effective
Hamiltonian, $H_{\text{eff}}^{(3)}$, it is apparent that the $E$ and
$F$ couplings are not invariant under time reversal, $T$, that is,
they change sign after complex conjugation. This leads to the
breaking of the chiral symmetry between the two opposite circulations
that atoms can take around a triangle.

By additionally taking $J^\downarrow=-J^\uparrow=J$, one can set all the couplings to be zero apart from $F$ and the effective
Hamiltonian reduces to
\begin{equation}
H_{\text{eff}}^{(3)}=F\sum_{\langle ijk
  \rangle}\vec{\sigma}_i\cdot \vec{\sigma}_j
\times \vec{\sigma}_k, \label{chiral}
\end{equation}
with $\vec{\sigma}=(\sigma^x,\sigma^y,\sigma^z)$ and
$F=|J|^3/U^2$. Remarkably, with this physical proposal, the
interaction term (\ref{chiral}) can be isolated, especially from the
Zeeman terms that are predominant in equivalent solid state
implementations. This interaction term is also known in the literature
as the {\em chirality operator} \cite{Wen}. It breaks the time reversal
symmetry of the system, a consequence of the externally applied field,
by effectively splitting the degeneracy of the ground state into two
orthogonal sectors, namely ``$+$'' and ``$-$'', related by time
reversal, $T$. These sectors are
uniquely described by the eigenstates of $H_{\text{eff}}^{(3)}$ at the
sites of one triangle. The lowest energy sector with eigenenergy
$E_+=-2\sqrt{3} F$ is given by
\begin{equation}
\begin{split}
&|\Psi^+_{1/2} \rangle = {1 \over \sqrt{3}} \big(|\uparrow
\uparrow \downarrow \rangle + \omega |\uparrow \downarrow \uparrow
\rangle + \omega^2 |\downarrow \uparrow
\uparrow \rangle \big) \\
& |\Psi^+_{-1/2} \rangle =- {1 \over \sqrt{3}} \big(|\downarrow
\downarrow \uparrow \rangle + \omega |\downarrow \uparrow
\downarrow \rangle + \omega^2 |\uparrow \downarrow \downarrow
\rangle \big) \label{states}
\end{split}
\end{equation}
where $\omega^3=1$. The excited sector, $|\Psi^-_{\pm1/2} \rangle$, represents counter
propagation with eigenvalue $E_-=2\sqrt{3}F$ and it is obtained from
(\ref{states}) by complex conjugation \cite{Wen,Rokhsar,Sen}. It has
been argued that this configuration, a result of frustration due to
the triangular lattice, and the disorder due to the presence of the
magnetic field, leads to analogous behaviour with the fractional
quantum Hall effect \cite{Wen,Kalmeyer} and, in particular, it can be
described by the $m=2$ Laughlin wavefunction defined on the lattice
sites. A recent example demonstrating this analogy is given in
\cite{Sorensen}.

\section{One- and Two-dimensional models} \label{onedim}

It is also possible to employ the three-spin interactions that we
studied in the previous sections for the construction
of extended one and two dimensional systems. The two dimensional
generalisation is rather straightforward as the triangular system
we considered is already defined on the plane. Hence, all the
interactions considered so far can be generalised for the case of
a two dimensional lattice where the summation runs through all the
lattice sites with each site having six neighbours.

\begin{center}
\begin{figure}[ht]
\resizebox{!}{2.5cm}
{\includegraphics{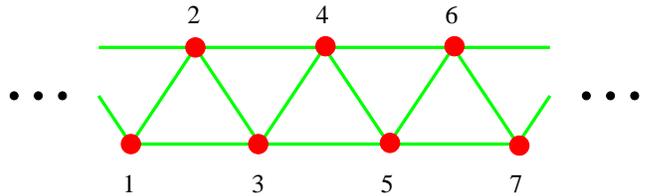} }
\caption{\label{chain} The one dimensional chain constructed out of
    equilateral triangles. Each triangle experiences the three-spin
    interactions presented in the previous sections.}
\end{figure}
\end{center}

The construction of the one dimensional model is more involved.
In particular, we now consider a whole chain of triangles in the
one dimensional pattern shown in Fig. \ref{chain}. In principle this
configuration can extend our model from the triangle to a chain.
Nevertheless, a careful consideration of the two spin interactions
shows that terms of the form $\sigma^z_i \sigma^z_{i+2}$
appear in the effective Hamiltonian, due to the triangular setting
(see Fig. \ref{chain}). Such
Hamiltonian terms involving nearest and next-to-nearest neighbour
interactions are of interest in their own right
\cite{Sachdev,Sachdev1} but will not be addressed here. It is
possible to introduce a longitudinal optical lattice
with half of the initial wavelength, and an appropriate amplitude
such that it cancels exactly those interactions generating, finally,
chains with only neighbouring couplings (for more details, see \ref{superlattice}).
In a similar fashion it is possible to avoid generation of
terms of the form $\sigma^x_i \sigma^x_{i+2} + \sigma^y_i
\sigma^y_{i+2}$ by deactivating the longitudinal tunneling coupling in
one of the modes, e.g. the $\uparrow$ mode, which deactivates the
corresponding exchange interaction.

As we are particularly interested in three-spin interactions we
would like to isolate the chain term $\sum_i (\sigma^x_i
\sigma^z_{i+1} \sigma^x_{i+2}+\sigma^y_i \sigma^z_{i+1}
\sigma^y_{i+2})$, the $\lambda^{(4)}$ term of Hamiltonian
(\ref{ham1}). This term includes, in addition, all the possible
triangular permutations. To achieve that, we could deactivate the
non-longitudinal tunneling for one of the two modes, e.g. the one
that traps the $\uparrow$ atoms. The interaction $\sigma^z_i
\sigma^z_{i+1} \sigma^z_{i+2}$ is homogeneous, hence it does not
pose such a problem when it is extended to the one dimensional
ladder. With the above procedures, we can finally obtain a chain
Hamiltonian as in (\ref{ham1}) where the summation runs up to the
total number, $N$, of lattice sites.

This class of Hamiltonians gives rise to a rich variety of ground
states for the spin ladder.  Let us consider, as an example, the case when
the tunnelling for one of the species is reduced to zero. This
switches off all terms involving $\sigma^x$ and $\sigma^y$ in the
Hamiltonian (\ref{ham1}). Suppose further that we have used additional
laser fields to cancel out the Zeeman term $B\sigma^z$.  Then, the
only non-zero coefficients in the Hamiltonian (\ref{ham1}) are
$\lambda^{(1)}$, the Ising interaction, and $\lambda^{(3)}$, the
3-spin interaction. The Hamiltonian becomes
\begin{equation}
H = \lambda^{(1)} \sum_i \sigma^z_i \sigma^z_{i+1}
  + \lambda^{(3)} \sum_i \sigma^z_i \sigma^z_{i+1} \sigma^z_{i+2}
\end{equation}

Without the 3-spin interaction, the Hamiltonian describes the
classical Ising model. The sign of the Ising interaction can be tuned
by the relative magnitudes of the repulsive potentials $U$.  For
negative $\lambda^{(1)}$, the system is ferromagnetic. The up-down
symmetry in the $z$-direction of the spin can be spontaneously broken
and there are two degenerate ground states: all spins up, or all spins
down.  For positive $\lambda^{(1)}$, the ground state is
the antiferromagnetic N\'eel state,
$|\uparrow\downarrow\uparrow\downarrow\uparrow\downarrow\cdots\rangle$
or
$|\downarrow\uparrow\downarrow\uparrow\downarrow\uparrow\cdots\rangle$. Note
that this breaks, spontaneously, the lattice translational symmetry with
a spin configuration of lattice period 2.

On the other hand, if we switched off the Ising interaction with a
suitable choice of $U$'s, the 3-spin term would give rise to different
broken-symmetry ground states. For $\lambda^{(3)}<0$, there are four
degenerate classical ground states: a ferromagnetic state
$|\uparrow\uparrow\uparrow\cdots\rangle$ and three states with lattice
period 3,
$|\uparrow\downarrow\downarrow\uparrow\downarrow\downarrow\cdots\rangle$,
$|\downarrow\uparrow\downarrow\downarrow\uparrow\downarrow\cdots\rangle$
and
$|\downarrow\downarrow\uparrow\downarrow\downarrow\uparrow\cdots\rangle$.
For $\lambda^{(3)}>0$, the ground states are the same with $\sigma^z$
reversed at each site.  Note that the 3-spin term explicitly breaks
spin-reversal symmetry, reflecting the difference in the
hopping amplitudes of the two atomic species of the original system.
Nevertheless, the lattice translational symmetry is spontaneously
broken in the period-3 states. A small Zeeman field in the
$z$-direction or a small antiferromagnetic Ising coupling will select
out the period-3 states from the ferromagnetic state.

We can switch directly between the period-2 and period-3 ground states
by tuning the interaction parameters $\lambda^{(1)}$ and
$\lambda^{(3)}$. The period-2 N\'eel state has an energy of
$-\lambda^{(1)}$ per site while the lowest-energy period-3 states have
an energy of $-\lambda^{(1)}/3 - |\lambda^{(3)}|$ per site.  So, we
expect a first-order phase transition along the line $2\lambda^{(1)}=
3|\lambda^{(3)}|$.

We can introduce quantum correlations into these models by introducing
a transverse Zeeman field, $B_x\sigma^x$, in the Hamiltonian at each
site
\begin{equation}
H = \sum_i \left[
  \lambda^{(1)} \sigma^z_i \sigma^z_{i+1}
  + \lambda^{(3)} \sigma^z_i \sigma^z_{i+1} \sigma^z_{i+2}
  + B^x \sigma^x_i \right]
\end{equation}
This transverse field gives rise to flipping between the up and
down spin states (of the $\sigma^z$ basis). The Ising chain (non-zero
$\lambda^{(1)}$ with $\lambda^{(3)}=0$) in a transverse field is a
well-known model. This is discussed extensively in \cite{Sachdev} and
it can be analytically solved using the Jordan-Wigner
transformation. As we increase the transverse field
$B_x$ from zero, the magnetisation in the $z$-direction (ferromagnetic
or antiferromagnetic according to the sign of $\lambda^{(1)}$) is
reduced. It vanishes at the quantum critical point
$B^x=|\lambda^{(1)}|$. At larger values of the transverse field, the
spins develops a polarisation in the $x$-direction. This quantum phase
transition for the one-dimensional chain belongs to the same
universality class as the two-dimensional classical Ising model with
no transverse field.

What happens when we introduce a weak 3-spin interaction into the Ising
chain with transverse field? Unfortunately, the model is not exactly
solvable. As already mentioned, the 3-spin term breaks spin-reversal
symmetry and so, just like a Zeeman field in the $z$-direction, it is
a relevant perturbation in the renormalisation group sense.

Indeed, considering the opposite limit where the 3-spin interaction is
strong, we know that a phase transition exists in the presence of a
transverse field with the 3-spin interaction but no 2-spin Ising
interaction \cite{Turb,Pens82}. At this transition, the period-3 order
parameter vanishes at $B^x= |\lambda^{(3)}|$.  Although this
transition looks analogous to the one found in an Ising chain in a
transverse field, it does not belong the same universality
class as the Ising chain. It is believed that it belongs to the
same universality class as the classical two-dimensional 4-state Potts
model \cite{Iglo87,Pens88}.

This leads us to speculate about how the direct transition between
period-2 and period-3 ground states is affected by the quantum
fluctuations introduced by a transverse field. From the viewpoint of
condensed matter physics, it will be interesting to study the critical
behaviour of this model. We believe that the critical point belongs to
the same universality class as the two-dimensional 3-state Potts
model, borrowing arguments described in \cite{huse82} for the melting
of commensurate structures.

Furthermore, we note that the phases of this spin chain are similar to
the quantum hard-rod system studied in \cite{Sachdev1}. That model has
the additional complexity of a macroscopically large number of
classically degenerate ground states at the classical transition
(without transverse field). We believe that our system is easier to
implement in the context of optical lattices for neutral atoms.

\begin{center}
\begin{figure}[ht]
\resizebox{!}{6cm}
{\includegraphics{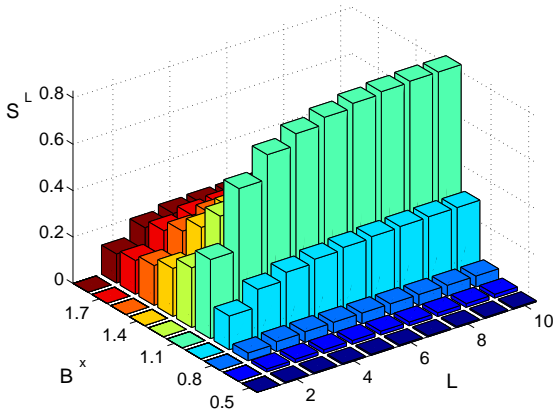} }
\caption{\label{entropy} The entropy $S_L$ as a function of the length
$L$ of the reduced density matrix and the transverse magnetic field
  $B_x$ for $\lambda^{(3)}=1$. The critical behaviour is indicated at
  the point $B_x \approx 1$. }
\end{figure}
\end{center}

From the viewpoint of quantum information, it is interesting to
study the degree of entanglement in this system. Indeed, we can
define a reduced density matrix for $L$ contiguous spins in a
system with $N$ spins by tracing out the other spins of the
system,
\[
\rho_L= {\rm tr}_{N-L} |\Psi_0\rangle\langle\Psi_0|
\]
where $|\Psi_0\rangle$ is the ground state of the system
\cite{Latorre04}. A measure of how these $L$ spins are entangled with
the rest of the chain is the von Neumann entropy,
\[
S_L = -{\rm tr} (\rho_L \log_2 \rho_L)
\]
We find that the entanglement between two halves of the system
increases dramatically near the critical point (Fig. \ref{entropy}),
similar to other studies of systems near criticality
\cite{Latorre04}. It will be interesting to investigate whether the
period-2 and period-3 states can be entangled by driving the system
through the transition.

\section{Localisable Entanglement and Cluster Hamiltonian}
\label{local}

In the past, Hamiltonians describing three-spin interactions have
been of limited interest \cite{Pens88,Iglo89} as they were
difficult to implement and control experimentally. The previous
results demonstrate that Hamiltonians with three-spin interactions
can be implemented and parameters controlled across a wide
intervals. One may suspect that ground states of three-spin
interaction Hamiltonians exhibit unique properties as compared to
ground states generated merely by two-spin interactions. This
motivates the study of the properties of the ground state of a
particular three-spin Hamiltonian for different parametric
regimes. Possible phase transitions induced by varying these
parameters are explored employing two possible signatures of
critical behaviour that are quite different in nature. In
particular, new critical phenomena in three-spin Hamiltonians that
cannot be detected on the level of classical correlations will be
demonstrated.

(i) A traditional approach to criticality of the ground state
studies two-point correlation functions between spins $1$ and $L$,
given by $ C_{1L}^{\alpha \beta} \equiv
    \langle \sigma_1 ^{\alpha} \sigma_L^{\beta} \rangle
    -\langle \sigma_1 ^{\alpha}\rangle
    \langle  \sigma_L^{\beta}\rangle
$, for varying $L$, where $\alpha,\beta=x,y,z$. These two-point
correlations may exhibit two types of generic behaviour, namely
(a) exponential decay in $L$, i.e. the correlation length $\xi$,
defined as
\begin{equation}
    \xi^{-1} \equiv \lim_{L\rightarrow\infty} \frac{1}{L}\log\,
    C_{1L}^{\alpha \beta},
    \label{corlength1}
\end{equation}
is finite or, (b), power-law decay in $L$, i.e. $C_{1L}^{\alpha
\beta}\sim L^{-q}$ for some $q$, which implies an infinite
correlation length $\xi$ indicating a critical point in the system
\cite{Sachdev}.

(ii) While the two-point correlation functions ${\cal
C}_{1L}^{\alpha \beta}$ are a possible indicator for critical
behaviour, they provide an incomplete view of the quantum
correlations between spins $1$ and $L$. Indeed, they ignore
correlations through all the other spins, by tracing them out.
Considering, for example, the GHZ state, $|GHZ\rangle = (|000\rangle
+ |111\rangle)/\sqrt{2}$, already shows that this loses important
information. Tracing out particle $2$ leaves particles $1$ and $3$
in an unentangled state. However, measuring the second particle in
the $\sigma_x$-eigenbasis leaves particles $1$ and $3$ in a
maximally entangled state. Therefore one may define the
localisable entanglement $E^{(loc)}_{1L}$ between spins $1$ and
$L$ as the largest average entanglement that can be obtained by
performing optimised local measurements on all the other spins
\cite{Verstraete PC 03}. In analogy to Eqn. (\ref{corlength1}) one
can define the entanglement length, $\xi_E$, by
\begin{equation}
    \xi_E^{-1} \equiv \lim_{L\rightarrow\infty} \frac{1}{L} \log\,
    E^{(loc)}_{1L}.
    \label{corlength2}
\end{equation}

\begin{center}
\begin{figure}[ht]
\resizebox{!}{6.5cm} {\includegraphics{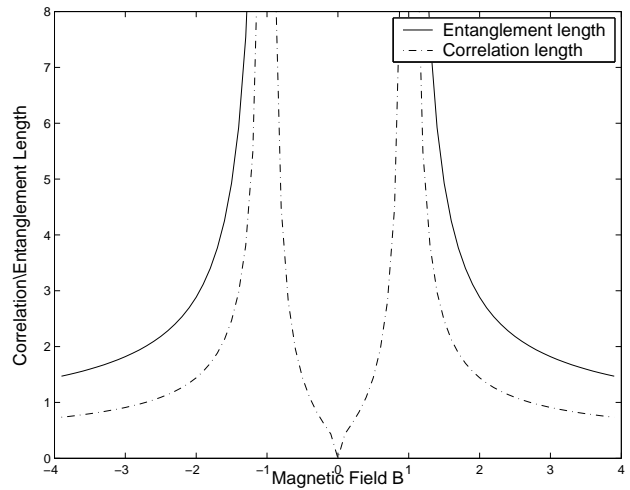} } \vspace*{0.2cm}
\caption{\label{entanglementlength} Both, the two-point
correlation length for $C^{zz}_{1L}$ (dashed line) and the
localisable entanglement length (solid line) are shown for various
magnetic field for chain of length 16. Note that the localisable
entanglement length diverges in the whole interval $|B|<1$ while
the two-point correlation length is finite in this interval. }
\end{figure}
\end{center}

It is now an interesting question whether criticality according to
one of these indicators implies criticality according to the
other. The localisable entanglement length is always larger than,
or equal to, the two-point correlation length and indeed, it has
been shown that there are cases where criticality behaviour can be
revealed only by the diverging localisable entanglement length
while the classical correlation length remains finite
\cite{Verstraete MC 03,Pachos04}. Such behaviour is also expected
to appear when we consider particular three-spin interaction
Hamiltonians. To see this consider the Hamiltonian
\begin{equation}
    H = \sum_{i}\big(\sigma^x_{i-1}\sigma^z_i\sigma^x_{i+1} +
    B\sigma^z_i \big),
    \label{xzxmodel}
\end{equation}
where we assume periodic boundary conditions. The fact that
$\sigma^x_{i-1}\sigma^z_i\sigma^x_{i+1}$ commute for different $i$
and employing raising operator $L^{\dagger}_k=\sigma^x_k -i
\sigma^x_{k-1} \sigma^y_{k}\sigma^x_{k+1}$ allows to determine the
entire spectrum of $H$ for $B=0$ quite easily. The unique ground
state of $H$ for $B=0$ is the well-known cluster state
\cite{Briegel R 99,Verstraete}, which has previously been studied
as a resource in the context of quantum computation. It possesses
a finite energy gap of $\Delta E=2$ above its ground state
\cite{comment}. For finite $B$ the energy eigenvalues of the
system can still be found using the Jordan-Wigner transformation
and a lengthy, but straightforward, calculation shows that the
energy gap persists for $|B|\neq 1$. The exact solution also shows
that the system has critical points for $|B|=1$ at which the
two-point correlation length and the entanglement length diverge.
For any other value of $B$ and in particular for $B=0$, the system
does not exhibit a diverging two-point correlation length as is
expected from the finite energy gap above the ground state.
Indeed, correlation functions such as
\begin{equation}
    \label{zzcorrelations}
    C_{ab}^{zz} = \psi_{ab}^2 - \chi_{ab}^2,
\end{equation}
with
\begin{equation}
  \psi_{ab}\equiv
  \frac{1}{4\pi}\int_{-2\pi}^{2\pi}
  \frac{\sin{r}}{\sqrt{B^2+1+2B\cos{r}}}
  \sin{\frac{(b-a)r}{2}}dr,
  \label{function1}
\end{equation}
and
\begin{equation}
  \chi_{ab}\equiv
  \frac{1}{4\pi}\int_{-2\pi}^{2\pi}
  \frac{B+\cos{r}}{\sqrt{B^2+1+2B\cos{r}}}
  \cos{\frac{(b-a)r}{2}}dr.
  \label{function2}
\end{equation}
can be computed and the corresponding correlation length can be
determined analytically using standard techniques (see
e.g. Fig. (\ref{entanglementlength})) \cite{Barouch M 71}. The
two-point correlation functions, such as Eqn.
(\ref{zzcorrelations}), exhibit a power-law decay at the critical
points $|B|=1$ while they decay exponentially for all other values
of $B$ in contrast to the anisotropic $XY$-model whose
$C^{xx}_{1L}$ correlation function tends to a finite constant in
the limit of $L\rightarrow\infty$ for $|B|<1$ \cite{Barouch M 71}.
This discrepancy is due to the finite energy gap the model in Eqn.
(\ref{xzxmodel}) exhibits above a non-degenerate ground state in
the interval $|B|<1$.

When we study three-spin interactions it is natural to consider
the behaviour of higher-order correlations. For the ground state
with magnetic field $B=0$, all three-point correlation except,
obviously,
$\langle\sigma^x_{i-1}\sigma^z_{i}\sigma^x_{i+1}\rangle$ vanish.
Indeed, if we consider $n>4$ neighbouring sites and choose for each
of these randomly one of the operators
$\sigma_x,\sigma_y,\sigma_z$ or $\mathbb{I}$ then the probability
that the resulting correlation will be non-vanishing is given by
$p=2^{-(2+n)}$. For $|B|>0$, however, far more correlations are
non-vanishing and the rate of non-vanishing correlations scales
approximately as $0.858^n$. This marked difference, which
distinguishes $B=0$, is due to the higher symmetry that the
Hamiltonian exhibits at that point.

In the following we shall consider the localisable entanglement
and the corresponding length as described in (ii). Compared to the
two-point correlations, the computation of the localisable
entanglement is considerably more involved due to the optimisation
process. Nevertheless, it is easy to show that the entanglement
length diverges for $B=0$. In that case the ground state of the
Hamiltonian (\ref{xzxmodel}) is a cluster state with the property
that any two spins can be made deterministically maximally
entangled by measuring the $\sigma_z$ operator on each spin in
between the target spins, while measuring the $\sigma_x$ operator
on the remaining spins. Indeed, this property underlies its
importance for quantum computation as it allows us to propagate a
quantum computation through the lattice via local measurements
\cite{Briegel R 99}.

For finite values of $B$ it is difficult to obtain the exact value
of the localisable entanglement. Nevertheless, to establish a
diverging entanglement length it is sufficient to provide lower
bounds that can be obtained by prescribing specific measurement
schemes. Indeed, for the ground state of (\ref{xzxmodel}) in the
interval $|B|<1$, consider two spins $1$ and $L=2k+1$ where $k\in
{\bf N}$. Measure the $\sigma_x$ operator on spin $2$ and on all
remaining spins, other than $1$ and $L$, measure the $\sigma_z$
operator.
By knowing the analytic form of the ground state one can obtain
the average entanglement over all possible measurement outcomes in
terms of the concurrence, that tends to $ E_{\infty} =
\left(1-|B|^2\right)^{1/4}$ for $k\rightarrow\infty$. This
demonstrates that the localisable entanglement length is infinite
in the full interval $|B|<1$. This surprising critical behaviour
for the whole interval $|B|<1$ is {\em not} evident from simple
two-point or n-point correlation function which exhibit finite
correlation lengths. For $|B|>1$ however, numerical results,
employing a simulated annealing technique to find the optimal
measurement for a chain of 16 spins, show that the localisable
entanglement exhibits a finite length scale.
From a practical point of view it is of interest to see how
resilient these predicted features are with respect to noise. To
this end, note that in Fig. (\ref{entanglementlength}) both the
two-point correlation length and localisable entanglement length
are drawn versus the magnetic field. In the interval $|B|<1$ the
entanglement length diverges while the correlation length remains
finite. For finite temperatures the localisable entanglement
becomes finite everywhere but, for temperatures that are much
smaller than the gap above the ground state, it remains
considerably larger than the classical correlation length. This
demonstrates the resilience of this phenomenon against thermal
perturbations.

The preceding analysis of our three-spin interaction Hamiltonian
(\ref{xzxmodel}) has indicated a marked qualitative discrepancy
between the localisable entanglement length and the two-point
correlation length with regard to variations in the magnetic field
strength parameter, $B$: across the interval between the two
critical points, $|B|<1$, the former quantity diverges while the
latter remains finite. We surmised that knowledge of simple
two-point correlation functions (such as $C_{ab}^{zz}$) alone is
insufficient in order to arrive at a complete characterisation of
criticality. Nevertheless, the question remains whether the
aforementioned discrepancy could not be reconciled by some
non-trivial {\it combination} (motivated by quantum information
theory) of two-point correlation functions. A natural and easily
computable candidate for such a quantity is the logarithmic
negativity \cite{logneg} (for basic ideas of the theory of
entanglement measures see, for example, \cite{EMeasures}), which
serves as a practical tool in the endeavour to quantify
multipartite entanglement. The logarithmic negativity of a quantum
state $\rho$ is defined as
\begin{equation}
    E_N \equiv \textrm{log}_2\tr |\rho^{\Gamma}|,
\end{equation}
where $\rho^{\Gamma}$ denotes the partial transpose of the density
operator and $tr|.|$ refers to the trace-norm, i.e. the sum of the
singular values of the operator.

We are interested in the entanglement, as measured by the
logarithmic negativity, between two spin-$\frac{1}{2}$ particles
residing at sites $i$ and $j$ on the chain modelled by Hamiltonian
(\ref{xzxmodel}). The composite state for the pair of spins is
described by the two-site reduced density matrix, $\rho_{ij}$,
which is formally obtained from the total density matrix
(describing the ground state of the whole chain) by tracing out
all the spin degrees of freedom apart from the two spins under
consideration.

In practice, however, the easiest procedure is to expand the
general two-site density matrix in the trace-orthogonal basis
formed by the tensor products of the Pauli spin operators at
either site as
\[
  \rho_{ij} =
  \frac{1}{4}\sum_{\alpha,\beta=0}^{3}
  \ev{\sigma_i^{\alpha}\sigma_j^{\beta}}
  \sigma_i^{\alpha}\otimes\sigma_j^{\beta},
\]
where the expansion coefficients are expectation values with
respect to the ground state of the Hamiltonian. It is useful to
represent this operator expansion in the standard two-qubit basis,
$\{\ket{\!\uparrow\uparrow},
\ket{\!\uparrow\downarrow},\ket{\!\downarrow\uparrow},
\ket{\!\downarrow\downarrow}\}$, where $\spinupket$ and
$\spindownket$ represent the eigenstates of $\sigma^z$ with
eigenvalues $+1$ and $-1$, respectively. In that basis,
\[
  \rho_{ij} =
  \sum_{\alpha,\beta=0}^{3}
  \ev{s_i^{\alpha}s_j^{\beta}}
  s_i^{\alpha}\otimes s_j^{\beta},
\]
with $s^0 \equiv\spinupket\!\spinupbra =
\frac{1}{2}\left(\mathbb{I}+\sigma^z \right)$, $s^1 \equiv
\spindownket\!\spindownbra = \frac{1}{2}\left(\mathbb{I} -
\sigma^z \right)$, $s^2 \equiv\spinupket\!\spindownbra =
\sigma^{+}$, $s^3 \equiv\spindownket\!\spinupbra = \sigma^{-}$.

A priori, the construction of $\rho_{ij}$ then requires knowledge
of sixteen expansion coefficients, \ev{s_i^{\alpha}s_j^{\beta}}.
Owing to symmetry properties of the Hamiltonian, however, these
are not all independent. For example, the Hamiltonian possesses
the global phase-flip symmetry $[H,U]=0$, where
$
  U = \otimes_{j=1}^{N}\sigma_j^z.
$
This symmetry then carries over directly to the two-site reduced
density matrix obtained from the global ground state in the form
$[\rho_{ij},\sigma_i^z\sigma_j^z]=0$, forcing half of its matrix
elements to vanish. Furthermore, translational invariance of the
Hamiltonian dictates that
$\rho_{\uparrow\downarrow,\uparrow\downarrow}
=\rho_{\downarrow\uparrow,\downarrow\uparrow}$ (here and
henceforth we drop the indices $i$ and $j$). Finally, the reduced
density matrix must be a real, symmetric matrix with unity trace.

Having taken the Hamiltonian's symmetries into consideration, we
find that the reduced density matrix possesses three distinct
diagonal elements,
\begin{align*}
  \rho_{1} & \equiv
  \rho_{\uparrow\uparrow,\uparrow\uparrow} =
  (1+2\ev{\sigma^z}+\ev{\sigma_i^z\sigma_j^z})/4,
  \\
  \rho_{2} & \equiv
  \rho_{\uparrow\downarrow,\uparrow\downarrow} =
  (1-\ev{\sigma_i^z\sigma_j^z})/4,
  \\
  \rho_{3} & \equiv
  \rho_{\downarrow\downarrow,\downarrow\downarrow} =
  (1-2\ev{\sigma^z}+\ev{\sigma_i^z\sigma_j^z})/4,
\end{align*}
and two distinct non-zero off-diagonal elements,
\begin{align*}
  \rho_{+} &=
  \rho_{\uparrow\uparrow,\downarrow\downarrow} =
  \ev{\sigma_i^{+}\sigma_j^{+}},\;\;\;\;
  \rho_{-} =
  \rho_{\uparrow\downarrow,\downarrow\uparrow} =
  \ev{\sigma_i^{+}\sigma_j^{-}}.
\end{align*}
The reduced density matrix is therefore uniquely determined by the
set of four expectation values,
\{\ev{\sigma^z},\ev{\sigma_i^z\sigma_j^z},
\ev{\sigma_i^{+}\sigma_j^{+}},\ev{\sigma_i^{+}\sigma_j^{-}}\}.

Since the reduced density matrix between nearest neighbours does
not exhibit entanglement for any value of the magnetic field $B$
we will now consider the reduced density matrix with respect to a
``bridge pair'' of spins (any pair of spins sandwiching a third).
Given the translational invariance of the Hamiltonian, we may
restrict our study to sites $1$ and $3$, without loss of
generality. The partial transpose of the reduced density matrix
with respect to either subsystem differs from the original matrix
only through an interchange of $\rho_{+}$ and $\rho_{-}$ (along
with \{i,j\} and \{1,3\}). Its four eigenvalues are given by
\begin{align*}
  \lambda_{1,2} &=
  \frac{(\rho_{1}+\rho_{3})
  \pm\sqrt{(\rho_{1}+\rho_{3})^2-4(\rho_{1}\rho_{3}-\rho_{-}^2)}}{2}
\end{align*}
and
\[
  \lambda_{3,4} = \rho_{2}\pm\rho_{+}.
\]

In order to quantify the entanglement within our bridge pair, all
that remains is to compute the set of four simple expectation
values that constitute the raw ingredients for the pair's reduced
density matrix. We are especially interested in the thermodynamic
limit, where $N \rightarrow\infty$ and we are no longer dealing
with complicated sums, but with manageable integrals. In fact, all
four expectation values are simple combinations of the two
integrals, Eqs. (\ref{function1}) and (\ref{function2}), defined
earlier:
\begin{align*}
  \langle\sigma^z \rangle &=-
  \chi_{00},\;\;\;\;
  \ev{\sigma_1^z \sigma_3^z} =
  \chi_{00}^2 + \psi_{13}^2 - \chi_{13}^2,
  \\
  \ev{\sigma_1^{+}\sigma_3^{+}} &=
  \frac{1}{2}\psi_{13}\ev{\sigma^z},\;\;\;\;
  \ev{\sigma_1^{+}\sigma_3^{-}} =
  \frac{1}{2}\chi_{13}\ev{\sigma^z}\, .
\end{align*}

For a few special values of $B$, these integrals, and the
associated logarithmic negativity, can be computed precisely with
ease. The results are summarised in the table below.
\begin{center}
\begin{tabular}{c|c|c|c|c|c|}
\cline{2-6}
&\multicolumn{5}{c|}{} \\[-8pt]
&\multicolumn{5}{c|}{{\bf magnetic field strength}}
\\[2pt]\cline{2-6}
&&&&& \\[-8.5pt]
&$-\infty$ &$-1$&$\phantom{..}0\phantom{..}$&$1$& $\infty$
\\[1.5pt]\hline
\multicolumn{1}{|c|}{}&&&&& \\[-8pt]
\multicolumn{1}{|c|}{$\psi_{13}$}
&0&$\frac{4}{3\pi}$&$\frac{1}{2}$&$\frac{4}{3\pi}$&\phantom{..}0\phantom{..}
\\[2pt]\hline
\multicolumn{1}{|c|}{}&&&&& \\[-8pt]
\multicolumn{1}{|c|}{$\chi_{13}$} &0&
$\frac{2}{3\pi}$&$\frac{1}{2}$&$\frac{2}{3\pi}$&0
\\[2pt]\hline
\multicolumn{1}{|c|}{}&&&&& \\[-8pt]
\multicolumn{1}{|c|}{$\langle\sigma^{z}\rangle$}
&$1$&$\frac{2}{\pi}$&$0$&$-\frac{2}{\pi}$&$-1$
\\[2pt]\hline
\multicolumn{1}{|c|}{}&&&&& \\[-8pt]
\multicolumn{1}{|c|}{$\langle\sigma_1^{z}\sigma_3^{z}\rangle$}
&$1$&$\frac{16}{3\pi^2}$ &$0$& $\frac{16}{3\pi^2}$&$1$
\\[2pt]\hline
\multicolumn{1}{|c|}{}&&&&& \\[-8pt]
\multicolumn{1}{|c|}{$\phantom{.}\langle\sigma_1^{+}\sigma_3^{+}\rangle$}
&$0$&$\frac{4}{3\pi^2}$&$0$&$-\frac{4}{3\pi^2}$&$0$
\\[2pt]\hline
\multicolumn{1}{|c|}{}&&&&& \\[-8pt]
\multicolumn{1}{|c|}{$\langle\sigma_1^{+}\sigma_3^{-}\rangle$}
&$0$&$\frac{2}{3\pi^2}$&$0$&$-\frac{2}{3\pi^2}$&$0$
\\[2pt]\hline
\multicolumn{1}{|c|}{}&&&&& \\[-8pt]
\multicolumn{1}{|c|}{$\sum_i|\lambda_i|$}
&1&$\frac{1}{2}+\frac{16}{3\pi^2}$&1&$\frac{1}{2}+\frac{16}{3\pi^2}$&1
\\[2pt]\hline
&&&&& \\[-8.5pt]
&0&\phantom{..}0.05711\ldots&0&\phantom{..}0.05711\ldots&0
\\[1.5pt]\cline{2-6}
& \multicolumn{5}{c|}{}\\[-8pt]
& \multicolumn{5}{c|}{{\bf logarithmic negativity}}
\\[2pt]\cline{2-6}
\end{tabular}
\end{center}

\begin{center}
\begin{figure}[h]
\resizebox{!}{6.5cm} {\includegraphics{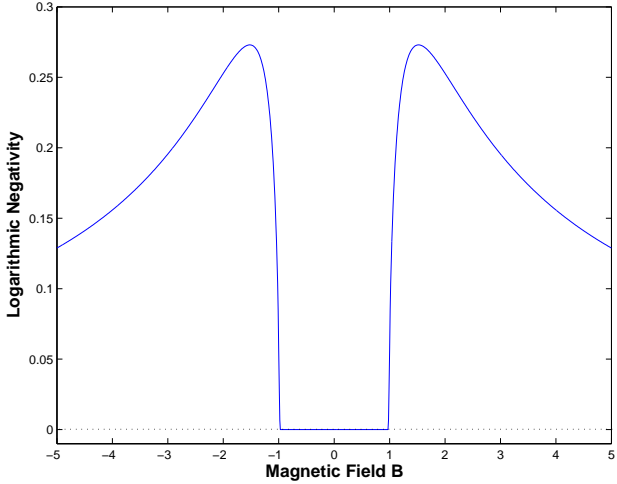} }
\caption{\label{lognegvsbfield} The logarithmic negativity is
shown as a function of the magnetic field strength parameter $B$
in the thermodynamic limit $N \rightarrow \infty$. }
\end{figure}
\end{center}

For general $B$, the integrals are most easily computed
numerically. A plot of the logarithmic negativity versus the
magnetic field strength is shown in Fig. (\ref{lognegvsbfield}).

\begin{center}
\begin{figure}[h]
\resizebox{!}{6.5cm} {\includegraphics{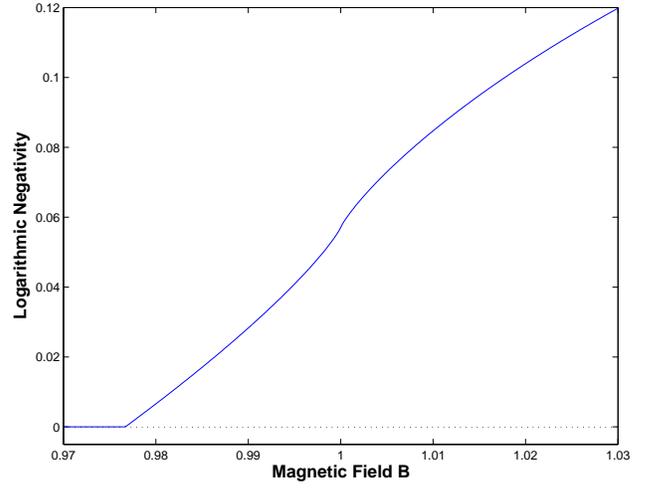} }
\caption{\label{lognegvsbfielddetail} Here, Fig.
(\ref{lognegvsbfield}) is magnified in a small interval centred
around the point $B_c=1$. The logarithmic negativity is clearly
seen to take on positive values before reaching $B_c=1$.}
\end{figure}
\end{center}

From among the features discernible in Fig.
(\ref{lognegvsbfield}), we are primarily interested in the
interval enclosed by the two critical points, $|B| \leqslant 1$,
where the logarithmic negativity appears to vanish identically.
That characteristic would be directly analogous to the diverging
localisable entanglement length encountered in Fig.
(\ref{entanglementlength}). Close inspection reveals, however,
that the logarithmic negativity grows positive even before
reaching the critical points, $B_c = \pm 1$. This phenomenon
already suggested itself as a result of our analytical computation
of the logarithmic negativity (c.f. the provided table), and is
most discernibly manifest in Fig. (\ref{lognegvsbfielddetail}),
which provides a close-up of the area immediately surrounding
$B_c=1$.

Note that we may bypass many steps involved in the computation of
the logarithmic negativity if we are not concerned with its
precise value across the whole range of $B$, but are merely
interested in establishing the region where it vanishes. For that
purpose, it suffices to check that the eigenvalues corresponding
to the reduced density matrix are all positive semi-definite. That
requirement is succinctly contained in the pair of inequalities
$\rho_{1}\rho_{3} \geqslant \rho_{-}^2$ and $\rho_{2} \geqslant
|\rho_{+}|$. In the event, it turns out that the latter inequality
is all that is needed here, because the former condition is
already satisfied for all values of the parameter $B$.

\begin{center}
\begin{figure}[h]
\resizebox{!}{6.5cm} {\includegraphics{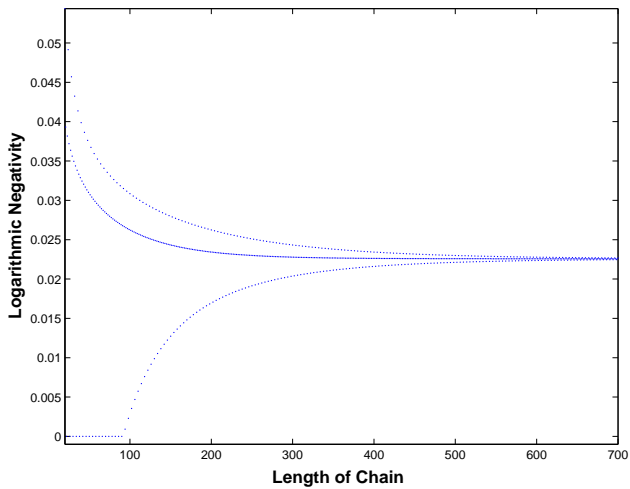} }
\caption{\label{lognegvslength} The logarithmic negativity with
respect to a bridge pair of spins is shown as a function of the
spin chain's length, with the magnetic field strength parameter
$B$ set to $0.9875$. The graph indicates a marked finite-size
effect, which vanishes in the thermodynamic limit. The middle
curve represents chains with an odd number of spins, while the
other two curves make up the even chain lengths. All three
curves converge to a common value as the chain length increases. }
\end{figure}
\end{center}

A note of caution with regard to predictions on the basis of spin
chains with finite length: such finite chains are prone to
uncharacteristic behaviour, particularly in the vicinity of a
quantum phase transition. As an example, let us consider how the
logarithmic negativity with respect to a bridge pair of spins
varies as a function of the spin chain's length. Such a scenario
is depicted in Fig. (\ref{lognegvslength}), with the parameter $B$
chosen to lie within the range of values where the logarithmic
negativity is non-vanishing in the thermodynamic limit (see Fig.
(\ref{lognegvsbfielddetail})); $B=0.9875$ to be precise.

Fig. (\ref{lognegvslength}) points to a remarkable finite-size
effect: the diagram is composed of three separate curves, the
middle one representing chains of odd lengths and the other two
together making up the even lengths. Furthermore, one of the
curves (representing lengths where $N+2$ is a multiple of four)
stays identically zero to start off with, until it abruptly rises
to converge with the other two curves. The diagram sees the
logarithmic negativity converge to a value around $0.0226$, which
is in complete agreement with the plot of Fig.
(\ref{lognegvsbfielddetail}) at the same value of $B$. For all
intents and purposes, $N=700$ can therefore already be regarded as
representing an infinitely long chain for this particular value of
$B$. Predictably, as the coupling parameter is steadily increased
to $B_c=1$, the three curves converge to an increasingly higher
value (reaching $0.5711...$ at $B_c=1$), and the rate of convergence
increases simultaneously. However, when we increase the coupling
parameter further still, another surprising finite-size effect
occurs. Precisely where the coupling parameter surpasses the
critical point $B_c=1$, the bottom curve of Fig.
(\ref{lognegvslength}) (now strictly positive for all $N$)
experiences a dramatic jump and approaches the convergence with
the other two curves {\it from above}. The size of this phenomenon
diminishes with increasing chain length and vanishes all together
for infinitely long chains, as is evident from Fig.
(\ref{lognegvsbfielddetail}). So it would actually appear that the
much sought-after critical behaviour in the entanglement measure
turns out to be a finite-size effect that vanishes in the
thermodynamic limit!

In summary, we embarked upon the preceding investigation in quest
of a simple order parameter that would allow us to predict the
occurrence of an infinite localisable entanglement length. A
combination of two-point correlation functions motivated by
quantum information science, the logarithmic negativity was deemed
to be a suitable candidate for such an order parameter, but our
findings have since led us to conclude that this measure cannot,
in general, predict the behaviour of the localisable entanglement
length. This further substantiates the belief that the localisable
entanglement length in translation invariant systems is a novel
concept that transcends mere two-point correlation functions.

\section{Quantum Computation} \label{qcomputation}

We have already seen how, in an optical lattice, for a
single triangle, one can manipulate, by varying suitably the
tunneling and/or the collisional couplings, the interaction terms
of the form $\sigma_j^x\sigma_{j+1}^x+\sigma_j^y\sigma_{j+1}^y$,
$\sigma_j^z\sigma_{j+1}^z$,
$\sigma_j^z\sigma_{j+1}^z\sigma_{j+2}^z$ and
$\sigma_j^z(\sigma_{j+1}^x\sigma_{j+2}^x+\sigma_{j+1}^y\sigma_{j+2}^y)$.
These interactions, up to common single qubit rotations, are
equivalent to quantum gates where the states $\ket{\uparrow}$ and
$\ket{\downarrow}$ represent the logical states $\ket{0}$ and
$\ket{1}$ of the computation. From the above interactions, one can
obtain SWAP, controlled-Phase ($CP$), controlled-controlled-Phase
($C^2P$) and controlled-SWAP (cSWAP) respectively. In the
following subsection we shall see how this is achieved. In
particular, the $CP$ and $C^2P$ gates are produced by manipulating
{\it one} of the two optical lattice modes while the other remains
with a high amplitude, corresponding to zero tunneling. The
exchange interaction (SWAP) is produced by lowering the barriers
that trap {\it both} the $\ket{0}$ and $\ket{1}$ states as seen in
(\ref{ham1}).

We shall then proceed to show how these gates can be combined to
perform quantum computation on one- or two-dimensional structures,
such as those described in section \ref{onedim}, without the need
for targeting specific lattice sites i.e. we will only control
fields that are applied globally, to the whole system
\cite{Benjamin:2000,Benjamin:2002a,Benjamin:2003b,Benjamin:2001b,Kay 04}.

There are two key ideas in targeting operations to specific qubits
by global addressing. Firstly, we employ a specific qubit as a
pointer and, secondly, we consider the effect of double wavelength
fields for addressing alternate triangles. We expand on both of
these ideas in subsequent subsections.

\subsection{Quantum Gates from Interactions}

In order to create quantum gates from the interactions that we have, let us
define
$$
\Lambda_i=\int_0^T\lambda^{(i)}dt
$$
where $\lambda^{(0)}=B$ (\ref{ham1}). This encapsulates the variation with
time of the coupling strengths as various optical lattices are applied. The
results should always be taken modulo $2\pi$, due to the exponentiation
procedure. The time, $T$, will be the
same for each $\Lambda_i$ for a given gate but can, naturally, be different
for the different gates that we wish to create.
\begin{table}[ht]
\begin{center}
\begin{tabular}{|c|c|c|c|c|c|}
\hline
Gate  & $\Lambda_0$ & $\Lambda_1$ & $\Lambda_2$ & $\Lambda_3$ & $\Lambda_4$
\\
\hline
$CP$  &$-\frac{\pi}{4}$ &$\frac{\pi}{4}$ & 0 & 0 & 0 \\
$C^2P$&$\frac{\pi}{8}$& $-\frac{\pi}{8}$ & 0 & $\frac{\pi}{8}$ & 0  \\
SWAP  & 0 & 0 & $\frac{\pi}{2}$ & 0 & 0 \\
cSWAP & 0 & 0 & $\frac{\pi}{4}$ & 0 & $\frac{-\pi}{4}$ \\
\hline
\end{tabular}
\label{tab_qgates}
\end{center}
\end{table}

Indeed, the above table shows how to create the quantum gates from the different
terms of the Hamiltonian (\ref{ham1}). As presented, both SWAP and
cSWAP generate an extra phase gate. This effect can also be negated, but has
been left out for simplicity.

\subsection{Organisation of the Computer}

\begin{center}
\begin{figure}[ht]
\resizebox{0.4\textwidth}{!}{\includegraphics{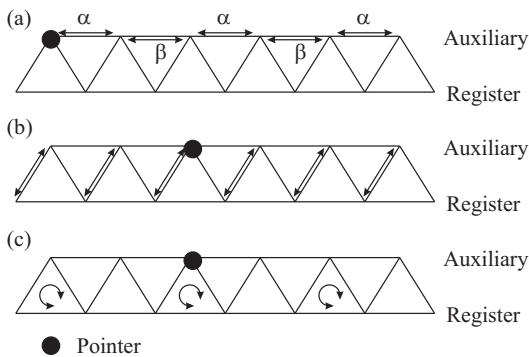}}
\caption{\label{computation}(a) The one dimensional chain of triangles
  divided into a row of auxiliary and register qubits. All are
  initialised as a $\ket{0}$ apart from the pointer, which is in the
  state $\ket{1}$. The interactions required to transport the pointer
  are also indicated. $\alpha$ and $\beta$ exchange interactions can be
  controlled separately. (b) Interactions required to create the $CP$
  gate, yielding $\sigma_z$ on a single qubit. (c) Interactions
  required to create $C^2P$ gate on alternate triangles.}
\end{figure}
\end{center}

We start with a one dimensional ladder of triangles, as shown in
Figure \ref{chain} and again in Figure \ref{computation}(a), although this can easily be extended to two
dimensions. We label one horizontal line as the register array, and
the other as the auxiliary array. The register array will contain the
qubits that we perform the computation on while the auxiliary array
will facilitate the transport of the pointer. The whole system is
initialised such that every qubit is in the $\ket{0}$ state. We
require a pointer to help us localise operations. This is achieved by
retaining single qubit control over a single lattice site, so we can
flip that qubit into the $\ket{1}$ state.

The basic idea is that if we can create a $CP$ gate between lattice
sites along the lines shown in Figure \ref{computation}(b), then this
acts as a $\sigma_z$ gate being applied only to the qubit adjacent to the
pointer. If we can also create a $C^2P$ gate on alternate triangles,
for example around the triangles shown in Figure
\ref{computation}(c), then this is just like creating a $CP$ gate
between the two qubits on the same triangle as the pointer without
affecting at all the rest of the register qubits. These
actions are the main building blocks for performing universal
quantum computation with our setup.

\subsection{Moving the pointer and qubits}

We have specified how certain gates can be created when the pointer is
next to a specific qubit, or when the pointer and two specific qubits
are all located at the vertices of the same triangle. However, we need
to know how to move the pointer and the qubits so that they can
interact as we want them to. To achieve this, we use the SWAP
operation applied to alternate qubits.

Moving the pointer is a relatively simple matter if we have a system
of SWAPs available. We just switch between the two SWAP-ing modes
along the top of the chain, denoted by $\alpha$ and $\beta$ in Figure
\ref{computation}(a). This allows us to move the pointer to any
arbitrary position on the chain. Exactly the same idea can be applied
to the register qubits, provided they are separated by an even number of
qubits. If they are separated by an odd number of qubits, then
applying SWAPs just causes the two qubits to move together, with
constant separation. To avoid this, we have to use the pointer to
apply a controlled-SWAP to one of the qubits so that it becomes
separated from the other qubit by an even number. We have already
shown one method for
generating this cSWAP, using the natural Hamiltonian of the system.
It can also be built by standard gates, which we shall see how
to construct in the following.

\subsection{Superlattices} \label{superlattice}

\begin{center}
\begin{figure}[ht]
\resizebox{0.4\textwidth}{!}{\includegraphics{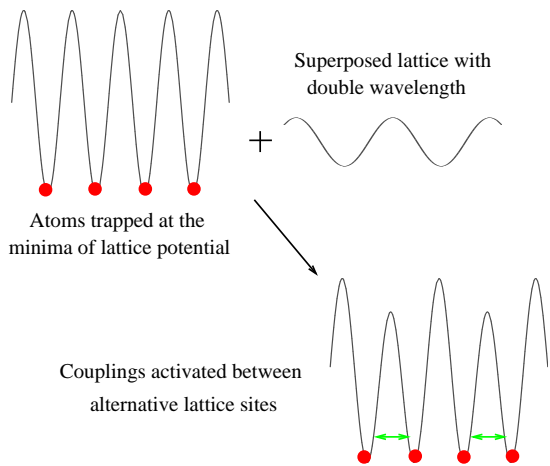}}
\caption{\label{super}Illustration of how a laser with double the
  wavelength of the trapping potential can activate alternate
  couplings, as required for the SWAP procedure (for example).}
\end{figure}
\end{center}

\begin{center}
\begin{figure}[ht]
\resizebox{0.4\textwidth}{!}{\includegraphics{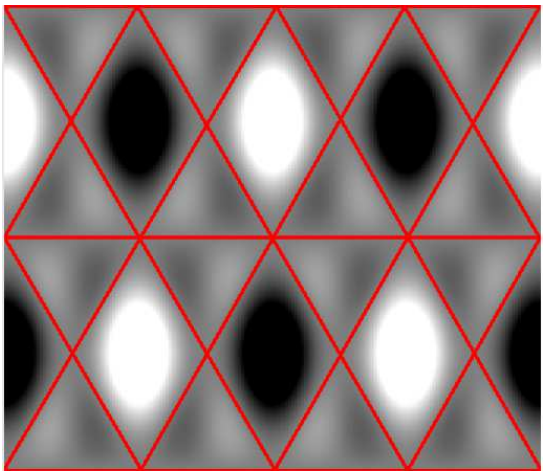}}
\caption{\label{potentialoffset} The potential offset that
activates tunneling transitions between alternate lattice sites.
The darker areas indicate lowering of the potential barrier, while
the lighter areas indicate its increase. The straight lines
indicate constant potential. The qubits are located at the vertices of
these lines.}
\end{figure}
\end{center}

In order to activate certain Hamiltonian terms on alternate triangles,
we need to employ the idea of superlattices. These are obtained by
superposing on the trapping potentials standing wave
fields with a different period. The idea is illustrated in Figure
\ref{super}. To generate these superlattices, it is not necessary to
have a large set of lasers with different wavelengths. Instead of
setting them up in direct opposition, it is possible to create
standing waves with varying periodicity by introducing an angle between
them \cite{Peil,Zoller:1}. This angle determines the period of
the standing wave, $d_i$ given by
$$
d_i=\frac{\lambda}{2\sin(\theta_i/2)}
$$
The required manipulation of potentials for triangular lattices
demands the activation of tunnelings along certain sites, while it
should be prohibited along other ones. For
example, to create the alternating SWAP that we require for moving the
pointer (Figure \ref{computation}(a)), we need to ensure that we don't
activate couplings between the register and the auxiliary arrays. We
also only want to activate couplings on every other horizontal
line. Hence, we specify that we require a potential of the form
$$
V_\text{off}= \cos(kx)\sin\left(\frac{ky}
{\sqrt{3}}\right)\sin\left
(\frac{ky}{\sqrt{3}}-kx\right)\sin\left(\frac{ky}{\sqrt{3}}+kx\right),
$$
taking the origin to be located on one of the register qubit lattice
sites. This can be expanded as a series of sine functions, each of
which can be created by a pair of lasers resulting in the
desired pattern, given in Fig. \ref{potentialoffset}. The determination of
orientation and angle for the laser pairs is demonstrated in
\cite{Kay 04} using the simpler example of a square lattice.

\subsection{Universal Quantum Computation}

So far, we have shown how to create a certain set of interactions as
we require them - $CP$, $C^2P$, SWAP and cSWAP. If we can also apply single
qubit rotations to alternate lines of the lattice, then we can apply
sequences such as $U$, $CP$ and $U^\dagger$, which have no effect if the
auxiliary qubit is in the $\ket{0}$ state because $U$ and $U^\dagger$
cancel, but applies $U\sigma_zU^\dagger$ to the qubit targeted by the
pointer. We are then able to create, by this process, any single qubit rotation on a
given qubit.  Together with the $CP$ gate (or
$\sqrt{\text{SWAP}}$ gate, which is generated using the same procedure
as for the $cSWAP$, for only half the time), it
constitutes a set of universal gates for quantum computation.

The single qubit rotations, created by Raman transitions on alternate
rows of the lattice, can, again, be generated by a standing wave of
double the period, so that the 0's of potential are localised on the rows that
we don't want to experience the rotation. The lasers that create this
standing wave need to be of a different wavelength to all the
others to avoid undesired transitions taking place.

\subsection{Error Prevention and Correction}

Errors are the bane of any implementation of quantum computation,
and there is much research on minimising their effects with better
quality apparatus and by using decoherence free subspaces
\cite{Ekert:96,Plenio 96}. It is known that once we are able to
perform operations with an error that is below a certain threshold
(assuming any errors leave the qubits within the computational
subspace), it is possible to perform arbitrarily accurate quantum
computation through the process of concatenation of quantum error
correcting codes \cite{Steane:96}, leading to fault tolerance
\cite{Steane:98}. It has recently been shown that the same can be
done for globally controlled systems such as the one presented
here \cite{Benjamin:2003b,
  Aharonov:99}.

The first thing to note is that the auxiliary array always remains in
a classical state. This instantly protects it against $\sigma_z$
errors, since classical states are eigenstates of $\sigma_z$. We can
also protect against $\sigma_x$ errors by
constantly measuring the state of the qubits via the Quantum Zeno
effect. Note that, if we measure the
auxiliary array in the $\ket{1}$ state, we risk losing the
pointer. We choose, instead, to measure only every other qubit on the
auxiliary array, not including the pointer. This at least provides an
indication of whether or not there has been an error on the pointer.

We can also employ this idea to provide measurement at the end of the
computation. We achieve
this by performing a cSWAP around a triangle so that we move a qubit
from the register array onto the auxiliary array, next to the
pointer. We are then able to perform the measurement.

Finally, the qubits on the register array can be encoded into a
quantum error correcting code. Sufficient parallelism can be
generated within the array to allow for error correction on every
block of encoded qubits (see \cite{Benjamin:2003b,Kay 04} for more
details). Instead of correcting the errors by making a measurement and
then performing the relevant correction, these steps are performed by
an algorithm that uses controlled-NOT gates to correct for the errors,
as determined by the error syndrome stored on ancilla qubits. The
problem with this method is that to perform the syndrome extraction
(placing on auxiliary qubits information about the errors), the
ancilla qubits must be in a well--known state ($\ket{0}$). After the
error correcting phase, these qubits are left in an unknown
state. Hence, it is necessary to either reset the state of the qubits,
or have a large supply of fresh qubits. In the optical lattice
set-up, this second option is quite sensible. We have so far restricted
computation to a single two-dimensional plane. However, in a three
dimensional lattice there are
many of these planes, all of which contain qubits in the $\ket{0}$
state. If we perform our computation on a single plane, then this
plane can be moved through all the other planes by a series of SWAPs
between alternate planes. We can therefore access this large supply of
fresh qubits for the purposes to error correction without the need
to perform measurements during the computation, thus simplifying
the experimental implementation.

\section{Conclusions}
\label{conclusions}

In this paper we presented, initially, a variety of different spin
interactions that can be generated by a system of ultra-cold atoms
superposed by optical lattices and initiated in the Mott insulator
phase. In particular, we have been interested in the simulation
and study of various three-spin interactions conveniently obtained
in a lattice with equilateral triangular structure. The
possibility to externally control most of the parameters of the
effective Hamiltonians at will renders our model as a unique
laboratory to study the relationship among exotic systems such as
chiral spin systems, fractional quantum Hall systems or systems
that exhibit high-$T_c$ superconductivity \cite{Wen,Laughlin}. 
Furthermore, unique properties related
with the critical behaviour of the chain with three-spin
interactions has been analysed (see also \cite{Pachos04}) where
the two-point correlations, used traditionally to describe the
criticality of a chain, seem to fail to identify long quantum
correlations, suitably expressed by a variety of entanglement
measures \cite{Verstraete PC 03}. In particular, analysing the
logarithmic negativity, indicates a possible connection between
the localisable entanglement length on the one hand and
entanglement properties of closely spaced spins on the other.
In addition, suitable applications have been presented within the
realm of quantum computation \cite{Pachos03,Kay 04} where
three-qubit gates can be straightforwardly generated from the
three-spin interactions. 

In conclusion the three-spin interactions generated in an optical
lattice offer a rich variety of applications in quantum
information technology as well as in solid state physics, worth
pursuing further theoretically and experimentally.

\acknowledgements

J. P. would like to thank Christian D'Cruz for useful conversations.
This work was supported by a Royal Society University Research
Fellowship, a Royal Society Leverhulme Trust Senior Research
Fellowship, the EPSRC QIP-IRC, the EU Thematic Network QUPRODIS
and by the Spanish grant MECD AP2001-1676. E.R. thanks the QI
group at DAMTP for their hospitality, where part of this work was
done.

\end{document}